\documentstyle[epsfig]{article}
\newcommand {\ds}   {\displaystyle} 
\newcommand {\be}   {\begin{equation}}
\newcommand {\ba}   {\begin{array}}
\newcommand {\bea}  {\begin{eqnarray}}
\newcommand {\bfi}  {\begin{figure}}
\newcommand {\ee}   {\end{equation}}
\newcommand {\ea}   {\end{array}}
\newcommand {\eea}  {\end{eqnarray}}
\newcommand {\efi}  {\end{figure}}                  
\newcommand {\dV}  {\delta V}
\newcommand {\UNIV}   {Universit\`a }
\begin{document}
%
\title{Double scaling and intermittency in shear dominated flows}
\author{
C\@.M\@. Casciola,
\thanks{Dip. Mecc. Aeron., \UNIV di Roma ``La Sapienza'',
        via Eudossiana 18, 00184, Roma, Italy.}
R\@. Benzi,
\thanks{ Dip. di Fisica and INFN, \UNIV di Roma ``Tor Vergata'',
         Via della Ricerca scientifica 1, 00133, Roma, Italy.}
P\@. Gualtieri$^*$,
B\@. Jacob$^*$,
\&
R\@. Piva$^*$.}
\maketitle
\section{Abstract}
The Refined Kolmogorov Similarity Hypothesis is a valuable tool for the description of 
intermittency in isotropic conditions. For flows in presence of a substantial mean shear, 
the nature of intermittency changes since the process of energy transfer is
affected by the turbulent kinetic energy production associated with the Reynolds stresses.
In these conditions a new form of refined similarity law has been found able to describe
the increased level of intermittency which characterizes shear dominated flows. 
Ideally a length scale associated with the mean shear separates the two ranges,
i.e. the classical Kolmogorov-like inertial range, below, and the shear dominated
range, above. However, the data analyzed in previous papers correspond to conditions 
where the two scaling regimes can only be observed individually.

In the present letter we give evidence of the coexistence of the two regimes and support 
the conjecture that the statistical properties of the dissipation field are 
practically insensible to the mean shear. This allows for a theoretical prediction of 
the scaling exponents of structure functions in the shear dominated range based on the 
known intermittency corrections for isotropic flows. The prediction is found to closely 
match the available numerical and experimental data.
\section{Introduction}
At large Reynolds number ($Re$) turbulent flows are characterized by strong non 
Gaussian intermittent fluctuations. For homogeneous isotropic turbulence, a quantitative
measure of intermittency can be given by using the structure functions 
$<\dV^p(r)>$ where
\be
\label{vel_diff}
       \dV(\vec r) = \left[\vec u(\vec x + \vec r) - \vec u(\vec x)\right] \cdot
       \frac{\vec r}{r}.
\ee
Then, the generalized dimensionless flatness 
\be
\label{flat_gen}
       F_p(r)=\frac{<\dV^p(r)>}{<\dV^2(r)>^{p/2}}
\ee
exhibits intermittency, in the sense that $F_p(r) \rightarrow \infty$ for 
$r \rightarrow 0$ and $Re \rightarrow \infty$. For $\eta \ll r \ll L_0$, where
$\eta$ is the Kolmogorov dissipation length and $L_0$ is the integral scale of 
turbulence, structure functions show scaling behavior, i.e. 
$<\dV^p(r)> \propto r^{\zeta(p)}$, where $\zeta(p)$ are anomalous scaling exponents 
($\zeta(p) \neq p / q \zeta(q)$), and $\zeta(3)=1$ due to the Kolmogorov 
{\em{four-fifth}} equation \cite{kolm_41}. It is a remarkable result, obtained in the 
last ten years, that $\zeta(p)$ are observed to be universal, i.e. independent of the 
Reynolds number and of the forcing mechanism, for homogeneous and isotropic turbulence 
\cite{benzi_1}.

Much less information is available for non isotropic turbulence. Recently, a number
of experimental \cite{anselmet} and numerical investigations \cite{PRL} in shear flow 
turbulence have shown that intermittency increases when the shear strongly affects 
the energy cascade. In the language of scaling exponents, an increase of intermittency 
means that $\zeta(p)$ are different from those observed in homogeneous and isotropic 
turbulence. Based on DNS of turbulent channel flow, it was recently proposed \cite{PF} 
that the increase of intermittency is due to the breaking of the Kolmogorov Refined 
Similarity Hypothesis (RKSH), which for homogeneous and isotropic turbulence reads 
\cite{kolm_62}
\be
\label{k62}
      <\dV^p(r)> \propto <\epsilon_r^{p/3}> r^{p/3}
\ee
where 
\be
\label{eps_r}
      \epsilon_r = \frac{1}{{\cal {B}}(r)} \int_{{\cal {B}}(r)} \epsilon_{loc} (\vec x)
                   d^3 x
\ee
and ${\cal {B}}(r)$ is a volume of characteristic size $r$ while 
$\epsilon_{loc} (\vec x)$ is the instantaneous local rate of energy dissipation. Equation
(\ref{k62}), in its Extended Self Similarity (ESS) formulation
\be
\label{rksh_old}
      <\dV^p(r)> \propto \frac{<\epsilon_r^{p/3}>}{<\epsilon_r>^{p/3}} 
                          <\dV^3>^{p/3}
\ee
has been successfully checked for a wide range of Reynolds numbers in different homogeneous
and isotropic turbulent flows \cite{benzi_2}.

For shear flow turbulence, it has been suggested that equation (\ref{rksh_old}) should
be replaced \cite{PF} by
\be
\label{rksh_new}
      <\dV^p(r)> \propto \frac{<\epsilon_r^{p/2}>}{<\epsilon_r>^{p/2}} 
                          <\dV^2>^{p/2}
\ee
for $r \gg L_s$ where $L_s=\sqrt{ \bar \epsilon / S^3}$, $S$ being the mean shear in the
system.  Following these considerations, the shear scale ideally separates, with regard
to the scaling properties of structure functions, the range of scales where turbulent
kinetic energy production prevails ($r \gg L_s$) from the range of scales characterized 
by purely inertial energy transfer ($ r \ll L_s$). Equation (\ref{rksh_new}) is able to explain 
the observed increase in intermittency, by assuming that the scaling properties of 
$<\epsilon_r^q>$ are not changed by the mean shear $S$.

It is the aim of this letter to present and discuss numerical and experimental evidence
that the two scaling regimes, predicted by eq. (\ref{rksh_old}) and
(\ref{rksh_new}), are indeed observed simultaneously in the range below and above $L_s$ 
respectively. Moreover, the results confirm the conjecture that the scaling properties of 
the energy dissipation are not modified appreciably with respect to homogeneous and isotropic 
turbulence.
\section{Data sets and analysis}
We discuss two sets of data, one obtained by a long and highly resolved DNS of homogeneous 
shear flow turbulence \cite{PF_new}, the other by hot wire measurements in the logarithmic 
region of a turbulent boundary layer \cite{jacob}.

Concerning the homogeneous shear flow, we have considered a turbulent flow with an 
imposed mean velocity gradient $S$ free from boundaries. The Navier-Stokes equations,
written in terms of velocity fluctuations, are solved by using an efficient pseudo-spectral 
method with a third order Runge-Kutta scheme for time advancement following the transformation
of variable proposed by Rogallo \cite{rogallo}. As shown by Pumir \cite{pumir} and recently 
confirmed by the present authors \cite{PF_new}, the flow reaches a statistical steady state 
characterized by large fluctuations of the turbulent kinetic energy.
The growth of turbulent kinetic energy is associated to large values of the Reynolds stresses,
produced by a well defined system of streamwise vortices via a lift-up mechanism \cite{kida}. 
In this flow, because of shear scale fluctuations due to the mentioned behavior of both
turbulent kinetic energy and Reynolds stresses, the crossover between the two scaling ranges 
is not sharply defined. In fact, we observe an overlapping of the two scaling regimes, and
the resulting scaling shows an effective slope.
In order to reduce as much as possible the 
fluctuations of the shear scale, a conditional sampling is introduced by 
considering only flow configurations where the production term exceeds a given threshold.
Among these configurations, only those corresponding to a large value of turbulent kinetic 
energy (${\cal{E}} > \alpha {\cal{E}}_{rms}$) are retained to reduce the mean value
of the ratio $L_s / \eta \propto \Omega^{3/4}$ ($\Omega$ being the mean enstrophy).

Concerning experiments, we analyze the velocity data on a flat plate boundary layer measured 
in a wind tunnel (test section length of $150 cm$) operated at $11.9 m/s$.
The boundary layer thickness is $\simeq 25 mm$ and the Reynolds number based on the momentum 
thickness is about $2200$. The boundary layer has the expected logarithmic region with the 
usual log-law constants \cite{monin}. Hot wire measurements were performed at several 
distances from the plate, using a constant temperature anemometer. The data acquisition was 
long enough to achieve convergence of the sixth order structure function.
\section{Double scaling regime}
We begin by analyzing the DNS of the homogeneous shear flow. We have strong evidence 
that for $r > L_s$, eq. (\ref{rksh_old}) fails and the new form of RKSH is 
established, as reported in \cite{cerci}, \cite{PF_new} and further discussed in the 
following. Furthermore,
the statistical properties of energy dissipation $<\epsilon_r^q>$ are not distinguishable 
from those observed in homogeneous and isotropic turbulence. The last statement can 
be directly checked by looking at figure \ref{fig_1} where we plot 
$<\epsilon_r^3>$ versus $<\epsilon_r^2>$ both for homogeneous shear flow and isotropic 
turbulence while, in the inset of the same figure, we plot $<\epsilon_r^2>$ versus 
$r / \eta$ for both cases. At variance with DNS, a direct measurement of
$<\epsilon_r^q>$ is not available for the experimental data and we are not fully confident 
in the one dimensional surrogate of $\epsilon_{loc}$ as a direct measure of the
local rate of energy dissipation, being the flow strongly anisotropic. Nevertheless by 
using the one dimensional surrogate, we can practically reproduce the 
results shown in figure \ref{fig_1}.

At any rate, to be cautious, we may avoid the explicit use of the energy dissipation
by plotting structure functions in the form suggested by Ruiz-Chavarria et. {\em{al.}}
\cite{ruiz}. Specifically, here, we introduce indicators based on $<\dV^p(r)>$ to 
detect the two scaling regions and to compare our findings with the predictions made in eq. 
(\ref{rksh_old}) and (\ref{rksh_new}). We remark that, both for numerical and experimental
data, the Reynolds number is not large enough to observe the scaling
of $<\dV^p(r)>$ and $<\epsilon_r^q>$ with respect to separation. Thus we 
employ the ESS as a valuable tool to estimate the scaling
exponents. In particular this implies that the exponents $\tau(q)$ are defined by the 
relation $<\epsilon_r^q> \propto <\dV^3>^{\tau(q)}$.

Following eq. (\ref{rksh_old}) and (\ref{rksh_new}) and the above discussion, we compute
both for the DNS and the experimental data the quantity 
$\sigma_p \equiv <\dV^p> / <\dV^2>^{p/2}$ and $\rho_p \equiv <\dV^p> / <\dV^3>^{p/3}$
which are expected to satisfy the relations:
\be
\label{ind_new}
      \sigma_p \propto 
      \left\{
      \ba{ll}
          \ds <\dV^3>^{\tau(p/2)} & \mbox{$r \gg L_s$} \\ \\
          \ds <\dV^3>^{\tau(p/3)- \tau(2/3) p/2} & \mbox{$r \ll L_s$}
      \ea
      \right.                                     
\ee
and
\be
\label{ind_old}
      \rho_p \propto 
      \left\{
      \ba{ll}
          \ds <\dV^3>^{\tau(p/2) - \tau(3/2) p/3} & \mbox{$r \gg L_s$} \\ \\
          \ds <\dV^3>^{\tau(p/3)} & \mbox{$r \ll L_s$}.
      \ea
      \right.                                     
\ee

Equations (\ref{ind_new}) and (\ref{ind_old}) allows us to compare the ESS exponents of
$\sigma_p$ and $\rho_p$ against the exponents predicted by equations (\ref{rksh_old}) and
(\ref{rksh_new}). Let us remark that eq. (\ref{ind_new}) and (\ref{ind_old}) are also
based on the assumption that $\tau(q)$ are the same both for shear dominated flows and
homogeneous and isotropic turbulence, as supported by the DNS data for 
$<\epsilon_r^q>$ shown in figure \ref{fig_1}.

In figure \ref{fig_2} we plot $\log \sigma_6$ against $\log <\dV^3>$ for the data
of the homogeneous shear flow and the turbulent boundary layer.
The fits for $r \gg L_s$ are in close agreement  with the value of 
$\tau(3)=-0.59$ expected from homogeneous and isotropic turbulence. In the inset 
of figure \ref{fig_2} we show the local slope $ d [\log \sigma_6] / d [\log <\dV^3>]$ 
computed from the homogeneous shear flow dataset. The two dashed lines indicate
the numbers $\tau(2)-3 \tau(2/3)$ and $\tau(3)$, i.e. the expected scaling exponents
for $r \ll L_s$ and $r \gg L_s$ respectively.

Figure \ref{fig_2} shows the main result of this letter, i.e. the clear evidence of the 
two scaling regions below and above $L_s$. Specifically, the experimental and numerical 
evidence of
the coexistence of two different intermittent regions in shear flow turbulence. As
a further check of the theory, in figure \ref{fig_3}, we show $\log \rho_6$ versus
$\log <\dV^3>$ both for the homogeneous shear flow and the turbulent boundary 
layer while in the inset we show the local slope 
$ d [\log \rho_6] / d [\log <\dV^3>]$. Also for the variable $\rho_6$ we can 
claim a very good agreement between the observed experimental and numerical results 
against theoretical predictions. Our results concerning the double scaling of 
structure functions are consistent and complementary with the generalized structure
function $<(\dV^3 + \alpha r \dV^2)^{p/3}>$ proposed by Toschi et. {\em{al.}} \cite{toschi}.

Finally we remark that, by denoting $\hat{\zeta}(p)$ the anomalous exponents in shear flow
turbulence and $\zeta(p)$ the anomalous exponent in homogeneous isotropic turbulence, 
the ESS estimate becomes
\be
\label{ESS_estimate}
       \hat{\zeta}(p) = \frac{p}{3} \left[ 1 - \tau\left(\frac{3}{2}\right) \right] 
                      - \frac{p}{2} + \zeta\left(\frac{3p}{2}\right).
\ee
The first term comes from the fact that in shear turbulence, we may express
\be
\label{rel_scaling}
       <\dV^3> \propto <\epsilon_r^{3/2}> <\dV^3>^{3 \zeta(2) / 2}
\ee
which implies $\hat{\zeta}(2) = 2/3 [ 1 - \tau(3/2)]$.
Equation (\ref{ESS_estimate}) provides a theoretical estimation for the scaling exponents 
of structure functions in shear dominated flows by using the intermittency corrections
of isotropic turbulence. The values given by eq. (\ref{ESS_estimate}) are compared against 
their direct measure in the DNS of the homogeneous shear flow in table \ref{tabella}. 
\section{Final remarks}
We have analyzed scaling properties of velocity fluctuations in shear dominated flows
considering a DNS of a homogeneous shear flow and hot wire measurements in the
logarithmic region of a turbulent boundary layer. When dealing with DNS data a 
conditional statistical analysis has allowed us to consider only shear dominated samples and
to reduce the fluctuations of the shear scale resulting in a well defined scaling
behavior below and above $L_s$ respectively. To by pass the use of the one dimensional 
surrogate of energy dissipation, in the experiments, we have evaluated the
refined similarity laws only in terms of structure functions. 

The analyzed data show clearly the coexistence of a double scaling behavior of structure 
functions across the shear scale. Actually the scaling exponents computed below and above 
the shear scale are in a very good agreement with the theoretical values provided by the 
classical RKSH \cite{kolm_62}, \cite{benzi_2}, and the new refined similarity law \cite{PF} 
respectively. Since the statistical properties of the energy dissipation are weakly affected 
by the shear, intermittency corrections of isotropic flows can be successfully employed in
the new form of similarity law, see table \ref{tabella}, to predict the scaling exponents 
of structure functions in the shear dominated range.
\newpage
\begin{table}
\caption{ Scaling exponents of structure functions (DNS) above and below the shear scale 
          $L_s$. Data are compared with those of homogeneous and isotropic 
          turbulence and with the prediction of eq. (\ref{ESS_estimate}).
          \label{tabella} }
\begin{tabular}{ccccccc}
\hline  \hline
 $p$                     & 1    & 2    & 3    & 4    & 5    & 6     \\ 
 $r < L_s$               & 0.36 & 0.69 & 1.00 & 1.28 & 1.54 & 1.78  \\ 
 $r > L_s$               & 0.38 & 0.72 & 1.00 & 1.23 & 1.42 & 1.56  \\
 $hom. iso$              & 0.36 & 0.69 & 1.00 & 1.28 & 1.54 & 1.78  \\
 eq. (\ref{ESS_estimate})& 0.39 & 0.73 & 1.00 & 1.23 & 1.42 & 1.58  \\
\hline  \hline
\end{tabular}
\end{table}

\bfi[h!]
\begin{center}
{\large{FIGURES}}
\end{center}
\vspace*{1.cm}
\centerline{
\epsfig{figure=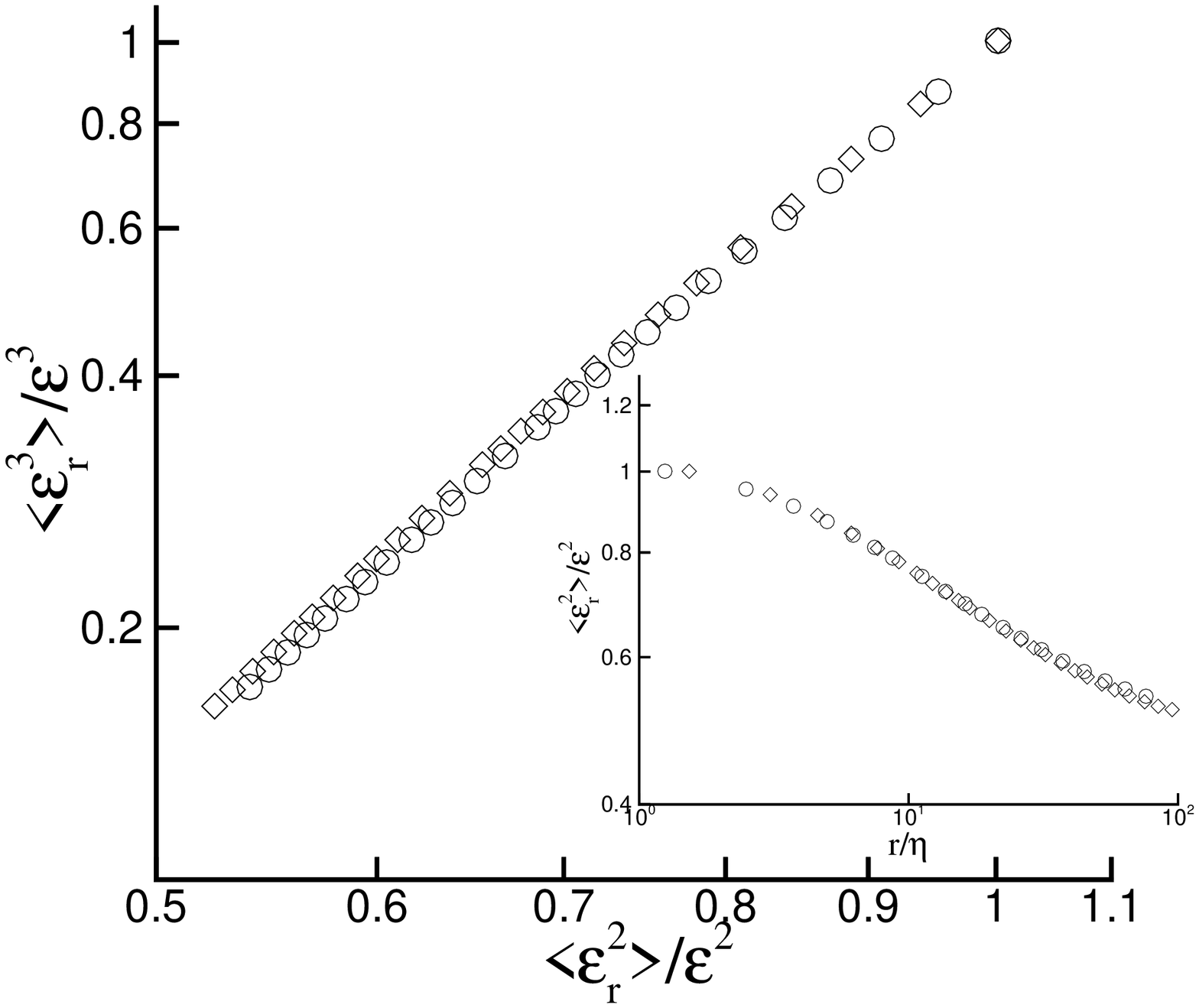,width=8.cm}}
\caption{$<\epsilon_r^3> \quad vs. \quad <\epsilon_r^2>$ in the homogeneous shear
         flow (circles) and in homogeneous isotropic turbulence (diamonds).
         In the inset $<\epsilon_r^2> \quad vs. \quad r / \eta$ for
         the two cases. \label{fig_1}}
\efi                                                                     
\newpage
\bfi[h!]
\centerline{
\epsfig{figure=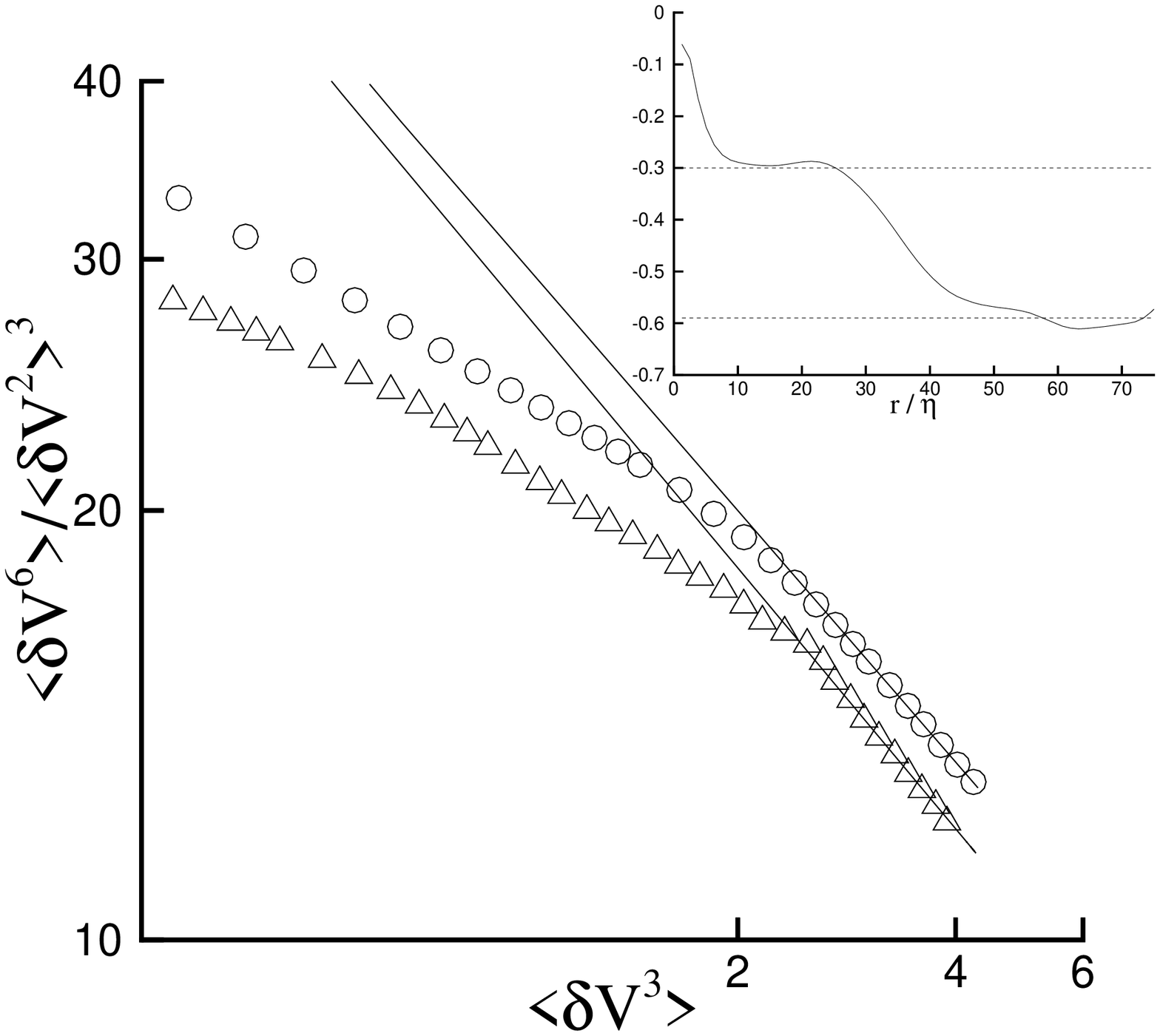,width=8.cm}}
\caption{$\log \sigma_6 \quad vs. \quad \log <\dV^3>$ in the homogeneous shear flow 
         (circles) and in the turbulent boundary layer at $y^+=115$ (triangles).
         DNS and experimental data are fitted at scales $ r > L_s$ by power laws
         with a slope $s=-0.58$ and $s=-0.59$ respectively. In the inset, the local slope
         $ d [\log \sigma_6] / d [\log <\dV^3>] \quad vs. \quad r / \eta$
         in the homogeneous shear flow obtained by considering the conditional sampling
         with $\alpha =1.3$ (solid line). The dotted lines corresponds to the
         two scalings given by eq. (\ref{ind_new}) at scales $r < L_s$ ($-0.3$) and
         $r > L_s$ ($-0.59$) using the values of $\tau(q)$ for isotropic
         turbulence. \label{fig_2}}
\efi                                                                     
\newpage
\bfi[h!]
\centerline{
\epsfig{figure=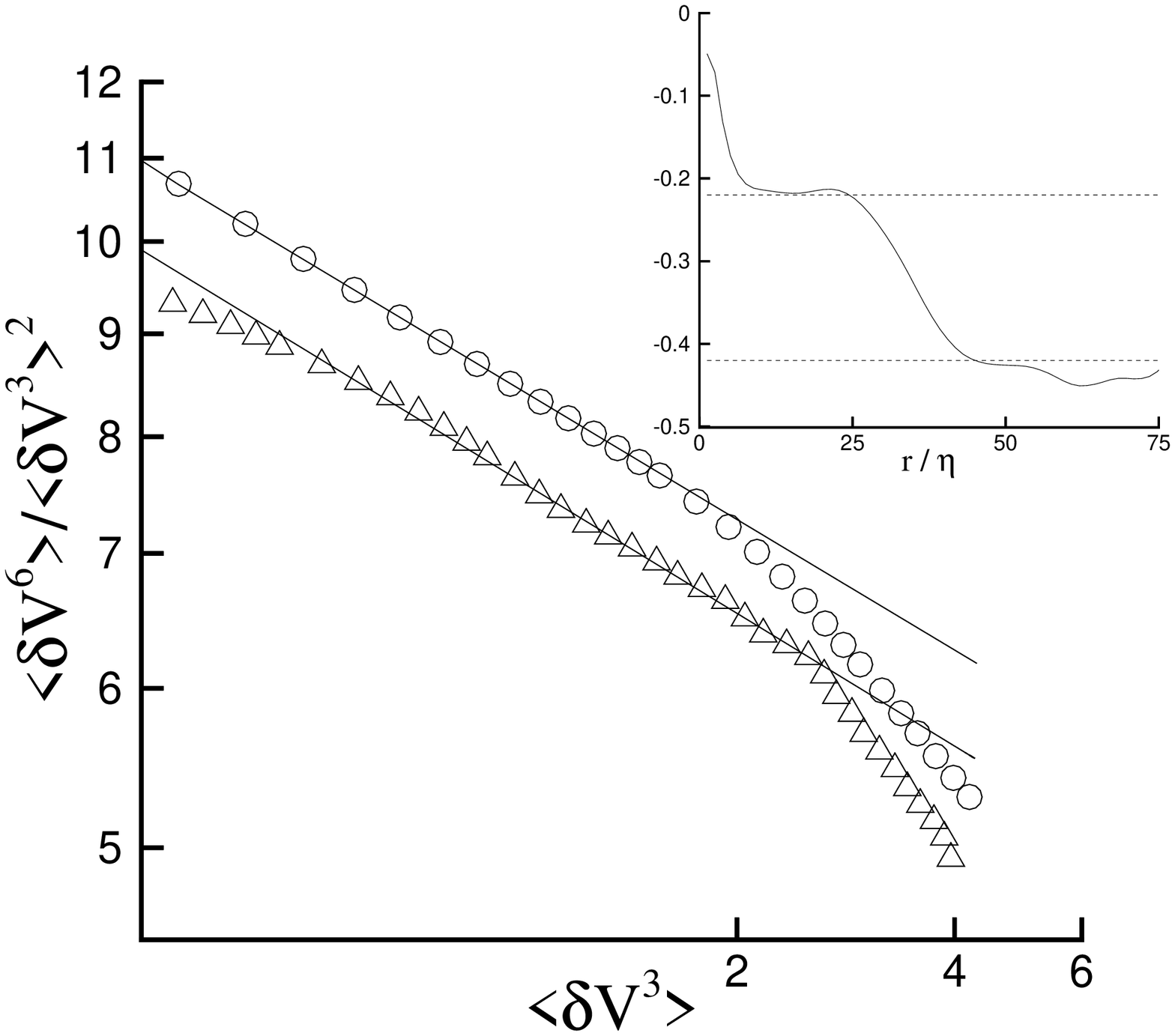,width=8.cm}}
\caption{$\log \rho_6 \quad vs. \quad \log <\dV^3>$ in the homogeneous shear flow
         (circles) and in the turbulent boundary layer at $y^+=115$ (triangles).
         Both DNS and experimental data are fitted at scales $ r < L_s$ by a power law
         with a slope $s=-0.22$. In the inset, the local slope 
         $ d [\log \rho_6] / d [\log <\dV^3>] \quad vs. \quad r / \eta$  
         for the homogeneous shear flow obtained by considering the conditional sampling 
         with $\alpha =1.3$ (solid line). The dotted lines corresponds to the
         two scaling given by eq. (\ref{ind_old}) at scales $r < L_s$ ($-0.22$) and
         $r > L_s$ ($-0.41$) using the values of $\tau(q)$ for isotropic
         turbulence.  \label{fig_3}}
\efi                                                                     
\end{document}